\documentclass[10pt]{article}
\usepackage{graphicx}
\usepackage{subcaption}
\usepackage[OE]{express}
\begin{document}
\title{Beam steering and impedance matching of plasmonic horn nanoantennas}

\author{Adeel Afridi,\authormark{1} \c{S}{\"{u}}kr{\"{u}} Ekin Kocaba\c{s},\authormark{1,*} }

\address{\authormark{1} Department of Electrical \& Electronics Engineering, Ko\c{c} University, Rumeli Feneri Yolu, TR34450 Sar{\i}yer, Istanbul, Turkey}

\email{\authormark{*}ekocabas@ku.edu.tr} 



\begin{abstract}
In this paper, we study a plasmonic horn nanoantenna on a metal-backed substrate. The horn nanoantenna structure consists of a two-wire transmission line (TWTL) flared at the end. We analyze the effect of the substrate thickness on the nanoantenna's radiation pattern, and demonstrate beam steering in a broad range of elevation angles. Furthermore, we analyze the effect of the ground plane on the impedance matching between the antenna and the TWTL, and observe that the ground plane increases the back reflection into the waveguide. To reduce the reflection, we develop a transmission line model to design an impedance matching section which leads to 99.75\% power transmission to the nanoantenna. 
\end{abstract}


%

\section{Introduction}
Optical antennas or nanoantennas have been the focus of research efforts due to their intriguing applications in wireless nano-links \cite{Yang2016,Yang2014,Merlo2016,Huang2009,Alu2010,Solis2013}, photodetection \cite{Knight2011}, photo emission \cite{Curto2010,Lee2011}, and nonlinear plasmonics \cite{Kauranen2012}, to name a few. Antennas play a vital role as the mediator between free space propagation and localized energy in communication systems \cite{Yang2016}. Antennas may offer less loss and improved performance as compared to their plasmonic waveguide interconnect counterparts \cite{Yang2016,Merlo2016,Alu2010}, specifically in optical wireless nano-links with long propagation distances of more than several wavelengths. However, modifications to the conventional antenna theory and design techniques are needed in the optical frequencies due to the changes in the properties of metals \cite{Novotny2007,Ramaccia2011}. 

Designing individual antennas to resonate at optical frequencies is not the only problem to be solved. Impedance matched integration of antennas with waveguides, efficient energy extraction from (or coupling to) the antennas, and controlling the directivity of the antennas are other challenges that need to be tackled \cite{Huang2009,Ramaccia2011,Xu2011,Sachkou2011,Ginzburg2007}. Recently, a widely used RF horn antenna \cite{Balanis2016} design has been extended to operate at the optical frequency regime. The horn geometry offers an inherently waveguide integrated design, a high directivity, and a large efficiency \cite{Yang2016,Yang2014,Ramaccia2011}. Ramaccia et al.\ first studied horn nanoantennas with a gradual exponentially tapered structure where Metal-Insulator-Metal waveguides with cylindrical metallic pillars were incorporated \cite{Ramaccia2011}. A similar structure was adopted by Yang et al.\ \cite{Yang2014} where they designed a broadband and highly directive E-plane horn \cite{Balanis2016} nanoantenna with straight flares fully incorporated in a homogeneous medium. In \cite{Yang2016}, Yang et al.\ extended the same horn antenna concept, where a plasmonic channel waveguide, along with the flared section, was carved in a silver film. Both \cite{Yang2016} and \cite{Yang2014} presented a systematic study of the horn nanoantenna for different flare parameters, where the antennas were embedded in a homogeneous medium, and radiated in end-fire direction\cite{Balanis2016} for wireless nano-link applications.

On the other hand, when a nanoantenna is placed at the interface between two different media with different refractive indices, it radiates a significant amount of power into the medium with the higher index \cite{Curto2010,Lee2011}. If horn nanoantennas presented in \cite{Yang2016,Yang2014}, which were designed for end-fire radiation, are placed at the interface of two heterogeneous media, they will radiate mostly into the substrate. As a result, only a fraction of the total power can be coupled or extracted from the end-fire direction, as the antenna radiation will be directed elsewhere. Thus, there is a need to engineer the radiation of these nanoantennas, so that they can radiate in any given direction, for applications in wireless nano-links \cite{Yang2016,Yang2014,Merlo2016,Huang2009,Alu2010,Solis2013}, and laser detection and ranging \cite{Sun2013}. Previously, plasmonic substrates (i.e.\ metallic ground-planes) were used to manipulate the nanoantennas' radiation pattern \cite{Xi2013, Min2011,Ahmed2011,Seok2011,Ghadarghadr2009}. If a reflector ground plane is employed, directly emitted light from the nanoantenna and the reflected light from the ground plane can interfere with different phases based on the thickness and type of the substrate material, enabling one to control and engineer the far-field radiation pattern \cite{Ghadarghadr2009,Xi2013}. In \cite{Ghadarghadr2009}, Ghadarghadr et al.\ reported that the dipole nanoantenna far-field radiation can be engineered by changing the distance between the dipole and the reflecting surface, and showed beam steering can be achieved in any arbitrary direction, ranging from end-fire to broadside. Similarly, reflector ground planes were also used by \cite{Min2011,Ahmed2011,Seok2011} in order to direct the scattered radiation from nanoparticles or dipole nanoantennas out of the substrate plane.

In this paper, we analyze and extend the plasmonic horn antenna design in \cite{Yang2014} to the case where the antenna is at the interface of air and a glass substrate with or without a metal mirror at its back. Analysis of the proposed design shows that beam steering can be achieved by changing the thickness of the substrate. In order to minimize the reflection from the antenna into the waveguide we use the impedance matching technique in \cite{Xu2011,Gagnon2008}. Rest of the paper is organized as follows. In Section \ref{sec:sec2}, we introduce the horn nanoantenna. Section \ref{sec:sec3} gives an in-depth analysis of the horn nanoantenna on an infinitely thick as well as on a finite thickness, metal-backed substrate, along with the illustration of impedance matching technique. Finally, we conclude in Section \ref{sec:sec4}.

\section{Antenna geometry}\label{sec:sec2}
\begin{figure}[!ht]
\centering 
\includegraphics[width=0.8\textwidth]{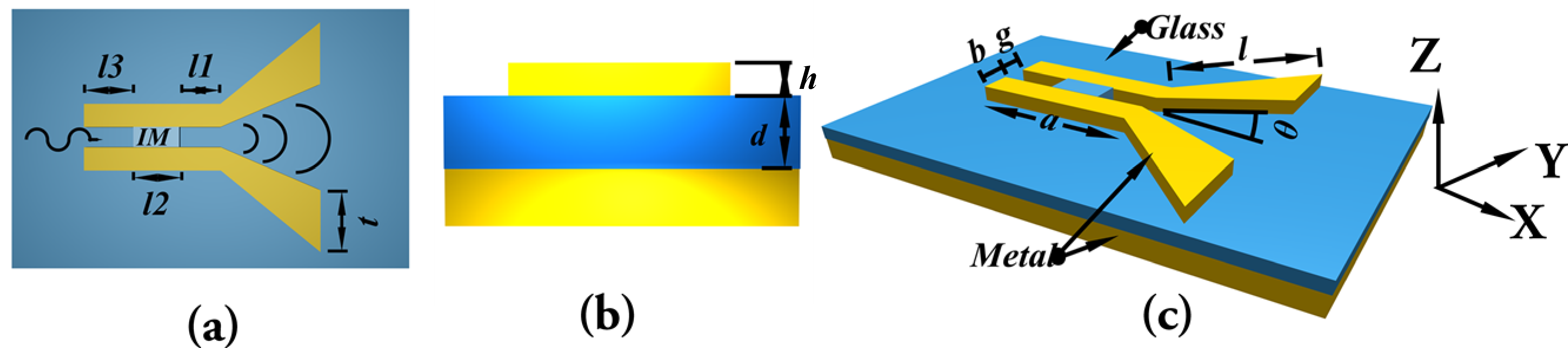}
\caption{Plasmonic horn nanoantenna. (a) Top view. (b) Side view. (c) 3D view.}
\label{fig1}
\end{figure}

The antenna structure with the ground plane and impedance matching section is shown in Fig.~\ref{fig1}. Figure~\ref{fig1}(a) shows top view of the antenna (XY-plane), and Fig.~\ref{fig1}(b) shows side view of the antenna (XZ-plane). The horn antenna, along with the waveguide, is placed at the interface between air (n=1) and glass (n=1.44). 
The waveguide is in the form of a two-wire transmission line (TWTL) with fixed dimensions $\{a,b,h\}$ given by $\{1000,100,50\}$ nm, respectively, while the gap $g$ between these two wires is equal to 30 nm. The horn length $l$ is 650 nm, the opening angle $\theta$ equal to 20$^{\circ}$, and width of the horn strip $t$ equal to 300 nm. It should be noted that lengths $l_1$, $l_2$, and $l_3$ make a total length $a$, where $l_1$ is position of the impedance matching (IM) section from flare and TWTL interface, and $l_2$ is length of the IM section. The substrate has a finite thickness $d$ in the case of a metal-backed antenna. The nanoantenna was simulated and analyzed in COMSOL MULTIPHYSICS. The whole structure was excited by the fundamental mode that is supported by the TWTL. A rectangular \textit{numeric port} was defined at the end cross-section of the TWTL, and \textit{boundary mode analysis} was used to calculate the mode at the specified port. The \textit{boundary mode analysis} performs an eigenfunction study to find the desired mode (fundamental mode in this case) supported by the TWTL, and its associated propagation constant. Consequently, the same mode, with input power ($P_\text{port} = 1\text{W}$) was fed to the antenna system. The whole structure was bounded by a box which served for the purpose of calculating the far-field pattern of the antenna, and a PML layer of finite thickness was assigned at each face of the bounding box. Permittivities for a range of wavelengths of the silver $\{Ag\}$ metal were obtained by interpolating the data from \cite{Johnson1972}.

\section{Results and discussion}\label{sec:sec3}
\begin{figure}[!ht]
\centering 
\includegraphics[width=0.9\textwidth]{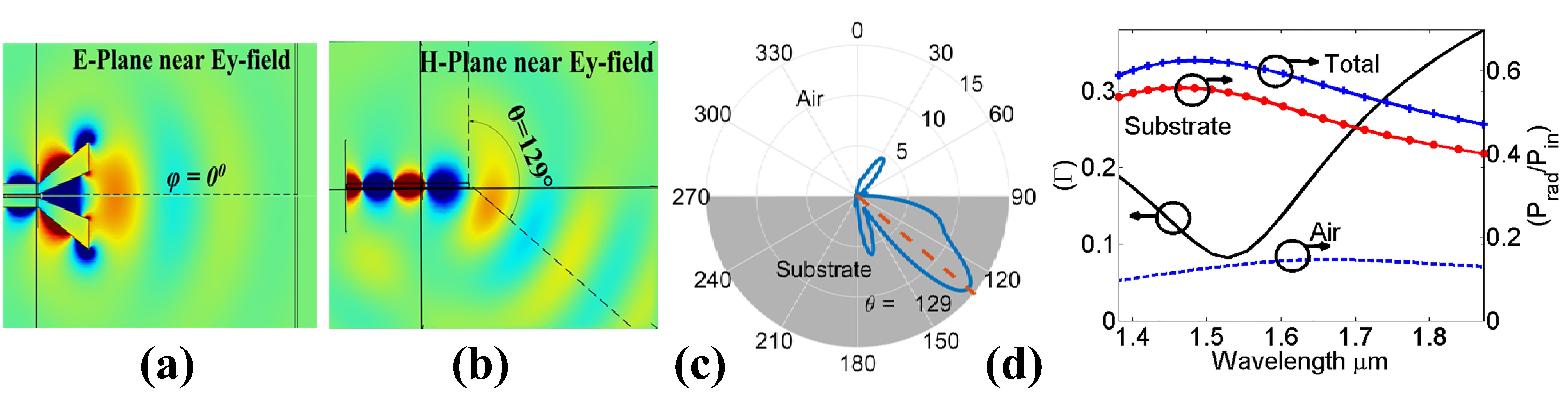}
\caption{Horn nanoantenna above an infinite substrate. (a) $E_y$ near field in the XY plane. (b) $E_y$ near field in the XZ plane. (c) H-plane directivity at 1550 nm. (d) Amplitude reflection coefficient $\Gamma$, and total radiation efficiency, radiation efficiency in the substrate, and in air.}
\label{fig2}
\end{figure}

In radio frequency (RF) and microwave engineering, an antenna's performance is evaluated by its near-field electric field (E-field) intensity, directivity, and how well the antenna is matched to the connecting waveguide by measuring the reflection coefficient $\Gamma$. Directivity is defined as $D(\theta,\phi) = 4\pi P(\theta,\phi)/ P_\text{rad}$, where $\theta$ and $\phi$ are elevation and azimuth angle, respectively, $ P(\theta,\phi)$ is the angular power density, and $P_\text{rad}$ is total radiated power. Figure~\ref{fig2} shows the results for the nanoantenna where the substrate is an infinitely thick glass. Figure~\ref{fig2}(a) and~\ref{fig2}(b) show the near-field $E_y$ intensity in the XY and XZ planes, respectively. It is seen in Fig.~\ref{fig2}(a) that the nanoantenna radiates, converting the bounded mode of TWTL to planar free space propagating waves by providing a gradual impedance match between TWTL and free space, while from Fig.~\ref{fig2}(b), one can observe that most of the radiation is directed into the substrate. Directivity of horn antennas is given in terms of E-plane and H-plane pattern \cite{Balanis2016}. E-plane is defined as the plane in which the primary E-field of the electromagnetic wave radiated by an antenna lies, which in our case is XY-plane, and H-plane is defined as the plane in which the magnetic field (H-field) lies, which is XZ-plane in this case. Figure~\ref{fig2}(c) shows the far-field H-plane directivity pattern of the antenna which has a maximum value of 14.5 at $\theta=129^\circ$. The amplitude reflection coefficient $\Gamma$ and the radiation efficiency $P_\text{rad}$/$P_\text{in}$ are given in Fig.~\ref{fig2}(d) as function of wavelength. At the central wavelength of 1550 nm, $\Gamma$ (black solid line) is less than 0.09 (in linear scale) meaning that $99.2\%$ power is delivered to the nanoantenna. During the calculation of the radiation efficiency, $P_\text{in}$  is calculated as $P_\text{in} = 10^{-a/L_p}$ \cite{Yang2016}, where $L_{p}$ is the waveguide's propagation length, and $a$ is length of the TWTL. The total radiation efficiency is around 0.61, at 1550 nm, while the efficiency in substrate is greater than efficiency in air, which also confirms that most of the power is radiated into the substrate.

\begin{figure}[!ht]
 \centering
    \includegraphics[width=0.9\textwidth]{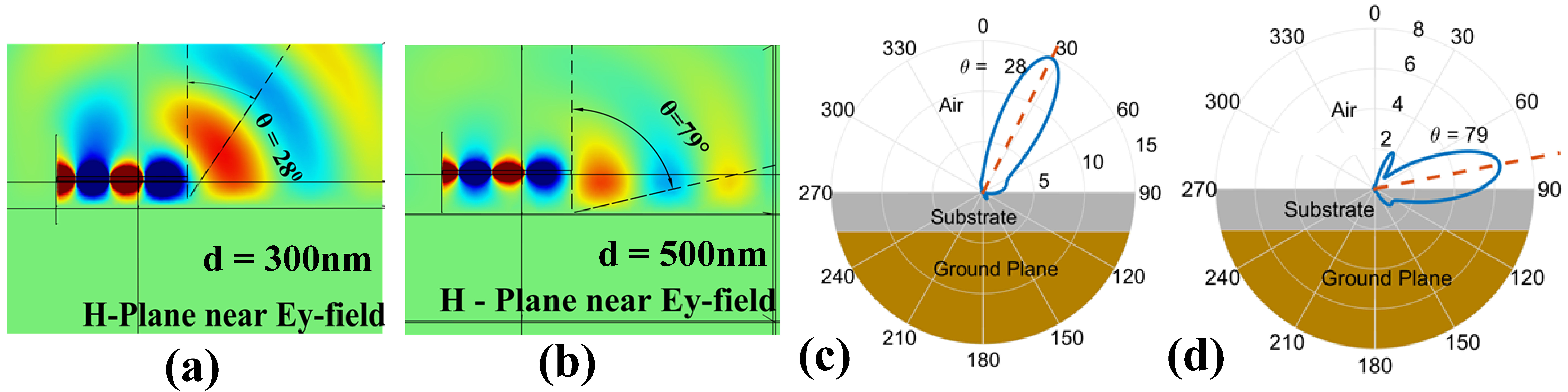}
    \caption{Results for the antenna with a ground plane. (a) $E_y$ near field in the XZ plane for $d$ = 300 nm. (b) $E_y$ near field in the XZ plane for $d$ = 500 nm. (c) \& (d) are corresponding H-plane directivities at 1550 nm.}
    \label{fig3}
\end{figure}

From our analysis of the nanoantenna with an infinite substrate, we see that most of the radiation is directed into the substrate. Therefore, only a fraction of the total power can be extracted from the plane parallel to the nanoantenna, as required in the case of a nano-link \cite{Yang2016,Yang2014}. In this section we investigate the effects of a reflective ground plane beneath the substrate. We examine the radiation properties by changing the thickness $d$ of the substrate. Figure~\ref{fig3} shows the results for a horn nanoantenna with a ground plane. Figure~\ref{fig3}(a) and~\ref{fig3}(b) show side view of near field $E_y$ of the nanoantenna for $d$ = 300 nm and 500 nm, respectively, where Fig.~\ref{fig3}(c) and~\ref{fig3}(d) show nanoantenna's directivity plot. For $d$ = 300 nm [Fig.~\ref{fig3}(a) and~\ref{fig3}(c)], antenna radiates most of the power into the air with a directivity of 15 at $\theta=28^\circ$. For $d$ = 500 nm [Fig.~\ref{fig3}(b) and~\ref{fig3}(d)], the antenna radiates end-fire, with a directivity of 7 at $\theta=79^\circ$. Similarly, the whole structure was simulated for a range of $d$ values between 200 and 900 nm, and we observed that for $d$ ranging from 200 nm to 450 nm, the antenna radiation was directed in air with the  maximum directivity at $\theta$ between $27^\circ$ to $40^\circ$. When $d$ was further increased above 450 nm until 650 nm, antenna radiated nearly end-fire with maximum directivity at $\theta$ ranging from $68^\circ$ to $79^\circ$. By further increasing $d$ above 650 nm again antenna radiated in air with almost same $\theta$ range as for below 450 nm. 

\begin{figure}[!ht]
 \centering
    \includegraphics[width=0.6\textwidth]{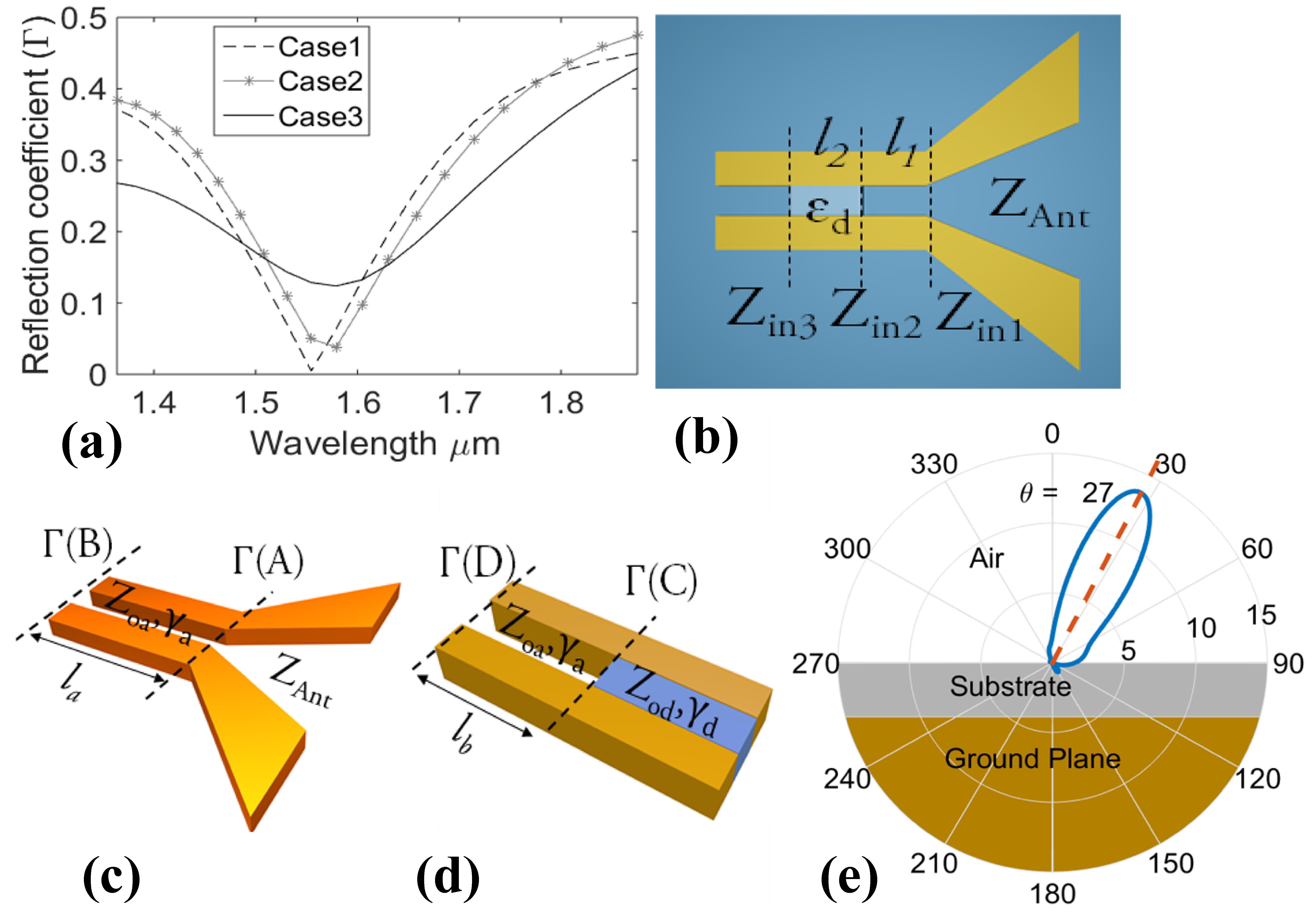}
    \caption{Antenna with ground plane ($d$ = 300 nm) and IM. (a) $\Gamma$ of: analytical model (case 1), simulated result (case 2) with IM, and simulated antenna without IM (case 3). (b) analytic model for IM. (c) air-core TWTL with antenna. (d) air-core TWTL cascaded with glass-core TWTL. (e) H-plane directivity at 1550 nm.}
    \label{fig5}
\end{figure}

The periodicity in antenna directivity can be explained by the antenna factor (AF) \cite{Balanis2016}, which is a sinusoidal function of thickness $d$ of the substrate, and the wave propagation constant. In addition, the periodicity of the radiation pattern is similar to the results reported in Figure 12(b) of \cite{Ghadarghadr2009}. The modal reflection coefficient $(\Gamma)$ plot for the antenna with $d = 300$ nm is given Fig.~\ref{fig5}(a) with a black solid line. As seen in the figure, $\Gamma$ increased to $\sim$0.13 from a value of 0.09 in the case of an infinitely thick substrate.

In order to reduce the back reflection from flares into TWTL, in this section we employ impedance matching (IM) scheme based on conjugate matching \cite{Xu2011}. For conjugate matching, a piece of dielectric (glass) was introduced in the gap of TWTL of the horn nanoantenna, shown in Fig.~\ref{fig1} and~\ref{fig5}(b), making a cascade matching network of total length $a$ equal to 1000 nm, comprised of air-core TWTL with length $l_3$, dielectric-core TWTL with length $l_2$, and again air-core TWTL with length $l_1$, from left to right as shown in Fig.~\ref{fig5}(b). For $\Gamma=0$, according to conjugate matching, impedance of the horn flares should be equal to the complex conjugate impedance of the matching network. Since impedance matching network in our case consists of the cascaded TWTLs with air or dielectric cores, input impedance at each interface [Fig.~\ref{fig5}(b)] can be obtained by transmission line equation as follows
\begin{equation}\label{eq:1}
	Z_{\text{in}N} = Z_{0N-1}[{1+\Gamma(N-1)\exp(-2\gamma_Nl_N)}]/[{1-\Gamma(N-1)\exp(-2\gamma_Nl_N)}].
\end{equation}
Here $Z_{0N-1}$ is the characteristic impedance of the TWTL \cite{Veronis2005} at the end of which interface N lies, $Z_{\text{in}N}$ and $\Gamma(N)$ are input impedance and reflection coefficient at interface N, and $\gamma_{N}$ and $l_N$ are the complex propagation constant and length of TWTL portion, respectively. Reflection coefficient $\Gamma$ is given by following equation
\begin{equation}\label{eq:2}
	\Gamma(N) = (Z_{\text{in}N} - Z_{\text{in}N+1})/ (Z_{\text{in}N} + Z_{\text{in}N+1}).
\end{equation}
Equation~\ref{eq:1} allows us to manipulate input impedance at each interface by changing $l_N$, and $\gamma_N$, and by input impedance manipulation one can minimize reflection coefficient $\Gamma$ according to Eq.~\ref{eq:2}. In order to achieve impedance matching, we propose an analytical model, which predicts optimal lengths $l_1$, and $l_2$ beforehand, rather than simulating for the range of lengths which is time and memory consuming. The analytical model is based on calculating normalized impedances at each interface of the horn nanoantenna with impedance matching network. Complete schematics of analytical model is given in Fig.~\ref{fig5}. It should be noted that all the structures used for modeling are placed on metal-backed substrate. We start our modeling by simulating the horn nanoantenna with air-core TWTL as shown in Fig.~\ref{fig5}(c). From the simulation, we extract the reflection coefficient $\Gamma(B)$ which is in turn used to calculate $\Gamma(A)$ by transforming it by $l_a$, according to equation $\Gamma(A)$ =$\Gamma(B)$$\exp(2\gamma_al_a)$. The normalized impedance $Z_{\text{Ant}}/Z_{0a}$ can be calculated by using Eq.~\ref{eq:2}. Similarly, as the second step of modeling, a cascade of air-core and dielectric-core TWTLs is simulated as shown in Fig.~\ref{fig5}(d), and $\Gamma(C)$ was calculated as $\Gamma(C)$=$\Gamma(D)$$\exp(2\gamma_al_b)$, where $l_b$, and $\gamma_a$ are length and propagation constant of air-core TWTL. Similarly, normalized impedance  $Z_{od}/Z_{0a}$ can be calculated again from Eq.~\ref{eq:2}. Finally, rest of the modeling includes calculation, where Fig.~\ref{fig5}(c) is cascaded to Fig.~\ref{fig5}(d), and forms Fig.~\ref{fig5}(b). Normalized input impedance at each interface can be calculated using Eq.~\ref{eq:1} and previous modeling steps. Thus, one can obtain impedance matching by changing the lengths $l_1$ and $l_2$. 

Figure~\ref{fig5}(a) and~\ref{fig5}(e) shows performance of antenna with ground plane and IM. Figure~\ref{fig5}(a) gives reflection coefficient comparison of theoretical model (case 1) with IM section, simulated result (case 2) with IM section, and simulated result without IM section (case 3). Optimal lengths, $l_{1}$ and $l_{2}$, equals to 350 nm and 125 nm respectively. 
\par
Theoretical model and simulated results are in agreement, thus allowing one to predict optimal lengths, and reflection coefficient of antenna system beforehand. Furthermore, IM reduced reflection of antenna from ${\sim}0.13$ to ${\sim}0.05$ (simulated case), which means that more than 61\% of improvement has been achieved. In addition, it can be observed from Fig.~\ref{fig5}(e) that IM does not have any significant effect on radiation pattern as the directivity of antenna remains the same.\par

\section{Conclusion}\label{sec:sec4}
In conclusion, the plasmonic horn nanoantenna with metal-backed substrate and IM has been investigated. We studied the nanoantenna placed on semi-infinite substrate and results demonstrate that most of the power is radiated into the substrate. We employ metal-backed substrate to achieve beam steering by varying substrate thickness. Results reveal that metal-backed substrate introduced impedance mismatch between the antenna and TWTL. We use impedance matching technique which allows us to accurately predict optimal parameters to minimize reflections. Impedance matching reduced reflections by more than 61\%, and 99.75\% of the input power is delivered to the nanoantenna. We use the nanoantenna in the transmission mode, converting surface plasmons to free space propagating waves. Conversely, the angular selectivity and the enhanced directivity of the horn nanoantenna can also be utilized in the receiving mode, and can be used for potential applications in wireless nano-links, sensing, and free space to chip communications.

\section*{Funding}
This work is supported by The Scientific and Technological Research Council of Turkey - TUBITAK (Project No. 112E247), TUBITAK - B{\.I}DEB 2215 - Graduate Scholarship Program for International Students, and Ko\c{c} University.
\end{document}